\documentstyle[12pt,draft]{article}
\newcommand{\beq}{\begin{equation}}
\newcommand{\eeq}{\end{equation}}
\newcommand{\beqa}{\begin{eqnarray}}
\newcommand{\eeqa}{\end{eqnarray}}
\newcommand{\la}{\lambda}
\newcommand{\ti}{\tilde}
\newcommand{\rh}{\rho}
\newcommand{\ga}{\gamma}
\newcommand{\da}{\dagger}
\newcommand{\al}{\alpha}
\newcommand{\si}{\sigma}

\newcommand{\om}{\omega}
\newcommand{\de }{\delta}

\newcommand{\LJ}{\langle}
\newcommand{\RJ}{\rangle}
\def\opone{\leavevmode\hbox{\small1\kern-3.8pt\normalsize1}}

\def\hO{\hat O}
\def\bO{\bar O}
\def\pt{\partial_t}
\def\intt{\int^t_0}
\def\half{\frac{1}{2}}
\def\e{{\rm e}}
\makeatletter                    
\@addtoreset{equation}{section}  
\makeatother

\begin{document}
\thispagestyle{empty}
\setcounter{page}{1}
\begin{titlepage}
\begin{flushright}
\end{flushright}
\vskip 1cm
\begin{center}
{\large{\bf Non-Markovian quantum state diffusion: Perturbation approach}}
\vskip 2cm 
{Ting Yu$^1$, Lajos Di\'{o}si$^2$,
Nicolas Gisin$^1$ and Walter T. Strunz$^3$\                                               
\vskip 0.4cm {\it
$^1$Group of Applied Physics, University of Geneva, 1211 Geneva 4,
Switzerland\\
$^2$Research Institute for Particle and Nuclear Physics,\\
1525 Budapest 114, P.O.B. 49, Hungary\\
$^2$Institute for Advanced Study, Wallotstrasse 19, D-14193 Berlin,
 Germany\\
$^3$Fachbereich Physik,Universit\"at GH Essen, 45117 Essen, Germany\\

}}
\vskip 0.7cm
\end{center}

\vskip 2.4cm

\begin{abstract}
We present a perturbation theory for non-Markovian quantum state 
diffusion (QSD), the theory of diffusive quantum trajectories
for open systems in a bosonic environment 
[Physical Review {\bf A 58}, 1699, (1998)].
We establish a systematic expansion in the ratio between
the environmental correlation time and the typical system time scale.
The leading order recovers the Markov theory,
so here we concentrate on the next-order correction corresponding to
first-order non-Markovian master equations. These perturbative 
equations greatly simplify the general non-Markovian QSD approach, 
and allow for efficient numerical simulations beyond the Markov
approximation. Furthermore, we show that each perturbative
scheme for QSD naturally gives rise to a perturbative scheme for 
the master equation which we study in some detail.
Analytical and numerical examples are presented, including
the quantum Brownian motion model.\\ 
{\bf PACS Numbers}: 03.65.-w, 03.65.Bz, 05.40.+j, 42.50.Lc
\end{abstract}

\end{titlepage}   
\section{Introduction}
Recently, a non-Markovian quantum trajectory theory - named
non-Markovian quantum state diffusion(QSD) - 
that describes the dynamics of a quantum `system' coupled
to an `environment' of harmonic oscillators
has been presented~\cite{DGS}.
Many outstanding new experimental advances can be properly studied only 
if non-Markovian effects are taken into account.
These include experiments with high-Q microwave cavities, quantum
optics in materials with a photonic bandgap, or output coupling from
a Bose-Einstein condensate to create an atom laser \cite{BLM,Ho,JQ,NJ,MHS},
to name a few. Also, the important phenomenon of decoherence
which takes place on time scales that can be of the same order as the 
correlation time of the environment require theories beyond the
standard Markov approximation. Further motivation are more 
fundamental questions about the proper description of individual open 
systems in quantum mechanics. Indeed the infamous problem of the
``Heisenberg cut" (understood here as the cut between the system and the
environment) is intimately related to the non-Markovian evolution of the 
system when the environment is ignored. 

In the Markov
regime, quantum trajectory approaches using stochastic Schr\"odinger
equations for pure states of the system play an important role in 
quantum optics, particularly for numerical simulations
\cite{DM,C,PK,GP,Perci,WiMi,CZA}.
These Markov models
also have many appealing features from a theoretical and conceptual 
point of view \cite{C,PK,GP,P1,DGHP,BYu}.
It is therefore desirable to generalize the powerful quantum
trajectory approach from the Markov regime to the more
general case of non-Markovian evolution. Several attempts
have been made recently from different perspectives. The linear 
non-Markovian unravelling underlying our theory was developed 
in \cite{D,S,DS} - see also~\cite{Pe} for a related attempt. 
Alternatively, a non-Markovian theory based on pseudomodes and a 
non-Markovian jump
approach have been developed recently \cite{Im,Ga,Ga1,BaLaMo,JCW}.

In ref. \cite{DGS}, the ultimate nonlinear non-Markovian QSD equation 
relevant to this paper has been derived directly from a microscopic 
model. In this framework, the reduced density matrix of the subsystem 
obtained by tracing over
the environmental degrees of freeedom
is unraveled into an ensemble of continuous trajectories
which correspond to the various realizations of the driving complex
Gaussian process. Non-Markovian QSD has been applied to interesting and 
physically relevant models where both computational power and many new 
features have been demonstrated~\cite{DGS}. However, many issues
are still to be addressed. In particular, further applications of
non-Markovian QSD to a variety of realistic problems are 
desirable. In addition, the theoretical implications of this new approach 
remain to be explored. Clearly, a full exploration of non-Markovian QSD
is an extensive project. The purpose of the present paper is a step 
towards this extensive project.

So far, although being a general theory,
it is difficult to implement non-Markovian QSD directly on a 
computer in all generality. It has been applied to a variety of
model problems in \cite{DGS}. However, in this paper we show how
the non-Markovian
QSD approach allows to find a systematic expansion of the
reduced system dynamics in powers of the ratio between the environmental 
correlation time and typical system time scales.
Thus, 
in order for non-Markovian QSD to have more 
applications, we establish a useful and practically relevant
perturbative approach that is directly amenable to computer simulations.
Our first and most important motivation for this paper is to present 
a systematic perturbative approach for non-Markovian QSD around the
Markov limit. This perturbative ``post-Markovian" QSD scheme is a
time-dependent
approach which preserves the non-Markovian nature of the problem in each
order of
approximation. 

The second purpose of the present paper is to establish
the relationship between the non-Markovian QSD equation and non-Markovian
master equations. Such a relation is well-known in the Markov regime,
where one generally starts from the standard Lindblad Markov master
equation to read off the Markov QSD equation \cite{PK,GP}. For
non-Markovian dynamics, closed master equations are rare
and thus, the direct link between non-Markovian QSD and the corresponding
master equation is of great interest. In order to obtain the master 
equation from
its non-Markovian QSD counterpart, one has to take the ensemble
mean over the stochastic process driving the QSD trajectories analytically 
(see also ref. \cite{DGS}). In general, this is only possible for 
simple models. In our pertubative scheme, however, we are able to derive 
master equations directly from non-Markovian QSD which turns out to be
useful from both theoretical and practical points of view.
It is important to note that a master equation derived in this way will 
necessarily preserve the positivity because such density
matrices can be decomposed into pure states at all times. 
We thus also address the problem how to ensure positivity for 
non-Markovian master equations - a difficult subject in itself.

The third motivation of the paper, therefore,
is to present a perturbation approach to non-Markovian
master equations.
Using an example we show that the resultant approximate master
equation preserves positivity. We also analyse the approximations leading 
to the Caldeira-Leggett model \cite{CaLe} - which is known to
violate positivity for certain initial conditions on short time scales. 

The organization of this paper is as follows. In Section II we briefly
review the basic concepts and equations of non-Markovian QSD. The aim of 
this section is two-fold: first, to act as a brief introduction for 
readers not familiar with the subject, second, to serve as the natural 
starting point of our new development. In Section III we develop the formal
time-dependent perturbation theory for non-Markovian QSD. In Section IV, we
present a systematic method of deriving the master equation from 
non-Markovian QSD.
We show that a perturbative QSD scheme naturally leads to a
perturbative scheme for the master equation.
We will apply the approximation schemes developed in this paper to some
examples in Section V.
In Section VI we take quantum Brownian motion as a typical model to
illustrate
the perturbative schemes based on QSD for the master equation. We conclude
the paper in Section VII, while some useful material can be found in the
Appendices.

\section{Non-Markovian quantum state diffusion}
Both Markov and non-Markovian QSD are based on two related stochastic
dynamical equations, a linear one and a nonlinear one. The linear one is 
mathematically simpler.  However, it does not preserve the norm of the 
state vector, which in general tends to zero.
Hence,
only the nonlinear equation, which preserves the norm, can be interpreted as
a distribution of time-dependent pure states with given probabilities (ie as
an unraveling): the
density matrix is then given by the ensemble mean of the pure states, at all
times. Moreover only this
nonlinear equation is suitable for numerical simulation \cite{DGS,DS}. 
Nevertheless, we start our 
presentation with the simpler linear equation, leaving the nonlinear one for
the following subsection.

\subsection{Linear non-Markovian quantum state diffusion}
\label{sec: lq}
Our quantum trajectory theory is based on a standard model of open
system dynamics: a quantum system interacting
with a bosonic environment with total Hamiltonian
\beq
H_{\rm tot} = H
+ \sum_\lambda g_\lambda (L a_\lambda^\dagger + L^\dagger a_\lambda)
+ \sum_\lambda \omega_\lambda a_\lambda^\dagger a_\lambda,
\label{Htot}
\eeq
where $H$ is the Hamiltonian of the system of interest and $L$, a system
operator coupling to environment, is called here Lindbald operator (as it 
plays the role of a `Lindblad operator' in the Markov limit). The 
linear non-Markovian 
QSD equation~\cite{D,S,DS} unravelling 
the reduced dynamics of model (\ref{Htot})
takes the form\footnote{For simplicity, we choose
$\hbar=1$ throughout the paper.}
\beq
\frac{d}{dt}\psi_t=-iH\psi_t + Lz_t\psi_t - L^\da\intt
\al(t,s)\frac{\de\psi_t}{\de z_s}ds,
\label{qsd}
\eeq
where $z_t$ is a colored complex Gaussian process with zero mean and
correlations
\beq
M[z_t^\ast z_s]=\al(t,s),\>\>\> M[z_tz_s]=0.
\eeq
The bath correlation function $\al(t,s)$ in (\ref{qsd}) has to be a 
positive  Hermitian kernel: $\al(t,s)=\al(s,t)^\ast$.
This non-Markovian unraveling is ensured to be consistent with
the reduced density operator approach since
the ensemble mean of the solutions of Eq. (\ref{qsd}) over
the noise $z_t$ will
reproduce the density matrix of the system,
\beq
\rh_t\equiv {\rm Tr_{\rm env}}\left(e^{-iH_{\rm tot}t}
\vert\psi_0\rangle\langle\psi_0\vert\otimes
\rho_0^{env}e^{iH_{\rm tot}t}\right)
= M\left[\vert\psi_t(z)\rangle\langle\psi_t(z)\vert\right].
\eeq
Here $M[\cdots]$ denotes the {\it ensemble average} over the classical
driving noise $z_t$.

From Eq. (\ref{qsd}), we see clearly that the evolution of the state
$\psi_t$ at $t$ depends on the whole history of the noise $z$. 
The equation
(\ref{qsd}) can
be written in the more appealing form
\beq
\frac{d}{dt}\psi_t=-iH\psi_t + Lz_t\psi_t - L^\da\intt
\al(t,s)\hat{O}(t,s,z)ds\psi_t
\label{qsdn}
\eeq
by defining an operator 
\footnote{ The notation
$\hat{O}(t,s,z)$,
rather than $\hat{O}(t,s,z_t)$ reflects that the operator $\hat{O}(t,s,z)$
contains the
noise $z$ in a nonlocal way, that is, it might be dependent on the whole
histories of
the noise $\{z_s: 0\leq s\leq t\}$.}
$\hat{O}(t,s,z)$ in (\ref{qsdn}) such that
\beq\label{ooperator}
\hat{O}(t,s,z)\psi_t \equiv \frac{\de\psi_t}{\de  z_s}.
\eeq
It turns out that $\hat{O}(s,s,z)=L$.
The $t$-dependence of the operator $\hat{O}(t,s,z)$ can be determined 
by the consistency condition
\beq
\frac{d}{dt}\frac{\de}{\de z_s}\psi_t = \frac{\de}{\de
 z_s}\frac{d}{dt}\psi_t
\label{consi}
\eeq
together with the linear non-Markovian QSD equation (\ref{qsdn}).

The appeal of equation (\ref{qsdn}) (or (\ref{qsd})) is its linearity
which is often useful in the mathematical analysis of our QSD approach
(see Section IV). Its use as simulation tool is severely undermined
since, for an infinite heat bath, the norm $\vert\vert\psi_t\vert\vert$ of
the solutions of (\ref{qsd}) goes to zero with probability one and to 
infinity with probability zero. 
For this reason, an unraveling in terms of normalized states is crucial 
for non-Markovian QSD to be truly useful for numerical simulations.

\subsection{Nonlinear non-Markovian quantum state diffusion}
\label{sec: nq}
The non-Markovian QSD unraveling based on normalized states
\beq
\tilde\psi_t(z) =\frac{\psi_t(z)}{\vert\vert\psi_t(z)\vert\vert}
\eeq
has been derived recently \cite{DGS} from the linear non-Markovian 
QSD equation (\ref{qsdn}) by making use of a Girsanov transformation
of the noise. We get
\beqa
\frac{d}{dt}\tilde\psi_t=&-&iH\tilde\psi_t +\left(L-\LJ L\RJ_t\right)
\tilde\psi_t\tilde z_t\nonumber\\
&-&\intt\al(t,s)\left[(L^\da-\LJ L^\da\RJ_t)\hat{O}(t,s,\tilde
z)\right.\nonumber\\
&-&\left.\LJ(L^\da-\LJ L^\da\RJ_t)\hat{O}(t,s,\tilde
z)\RJ_t\right]ds\tilde\psi_t
\label{nnm}
\eeqa
where $\tilde z_t$ is the shifted noise,
\beq
\tilde z_t =z_t + \intt\al(t,s)^\ast\LJ L^\da\RJ_s ds,
\eeq
and $\LJ L\RJ_t\equiv \LJ\tilde\psi_t\vert L\vert \tilde\psi_t\RJ$ denotes
the {\it quantum average}. Again here,
the {\it ensemble average} of the solution to Eq. (\ref{nnm}) recovers
the density matrix of
the system,
\beq
\rh_t = M\left[\vert\ti\psi_t(z)\rangle\langle\ti\psi_t(z)\vert\right].
\eeq
This nonlinear non-Markovian QSD equation can be rewritten in a more
compact form:
\beq
\frac{d}{dt}\ti\psi_t=-iH\ti\psi_t +\Delta_t(L)\ti\psi_t\ti z_t
-\Delta_t(L^\da)\bar{O}(t,\ti z)\ti\psi_t +
\LJ\Delta_t(L^\da)\bar{O}(t,\ti z)\RJ_t\tilde\psi_t
\label{nnm1}
\eeq 
where $\Delta_t(A)\equiv A-\LJ A\RJ_t$ for any operator $A$
and 
\beq
\bar{O}(t,z)=\intt\al(t,s)\hat{O}(t,s,z)ds.
\label{nota}
\eeq

Equation (\ref{nnm}) (or (\ref{nnm1})) is the fundamental equation of
non-Markovian QSD
\footnote{From now on, unless otherwise emphasized,
non-Markovian QSD refers to equation (\ref{nnm}) or (\ref{nnm1}), not 
to the linear equation (\ref{qsd}).}, and is our starting
point for the perturbative approach in Section III. Numerical
simulations of non-Markovian open system dynamics
using this equation can be found in ref. \cite{DGS}. Note
that the non-Markovian QSD equation (\ref{nnm}) reduced to the standard 
Markov QSD equation \cite{GP} for $\alpha(t,s)\rightarrow\delta(t-s)$.

To conclude this section, we would like to make two
remarks about the non-Markovian QSD approach.
First, the derivation of both the linear and
non-linear non-Markovian QSD equations are based on the assumptions that
the environment is bosonic, and that initially
the state of the total system+environment is
factorable $\rh_0=\rh_0^S\otimes \rh_0^{env}$,
where the  initial state of the system
$\rh_0^S=\vert\psi_0\RJ\LJ\psi_0\vert$ is
independent of the noise $z_t$. In fact, if $L\neq L^\dagger$, 
(\ref{nnm}) is valid for a zero temperature environment
only and the equation for
finite temperature gets two additional terms, see \cite{DGS}.
Secondly, since Eq. (\ref{qsd}) and hence
(\ref{nnm}) are derived directly from the microscopic model,
these QSD equations can be read
off automatically from the total Hamiltonian (\ref{Htot}).
This suggests that the non-Markovian QSD approach also represents a 
brand new way to derive the quantum master equation of
open quantum systems (see Section IV).

\section{Time-dependent perturbation theory}
\label{sec: per}
The non-Markovian QSD approach offers a very promising method
to handle
quantum systems whenever non-Markovian effects are relevant.
However, many interesting problems which arise in open quantum systems
are such that the operator $\hat{O}(t,s,z)$ appearing in Eq (\ref{nnm})
cannot be determined exactly. Moreover, the nonlocal noise contained in
the fundamental
equation (\ref{nnm1}) might cause difficulties in numerical
simulations.
In this section, we aim for a formal perturbation scheme for the
non-Markovian QSD
equation (\ref{nnm}). Applications of the general
perturbative
method developed here will be presented in Section V.

\subsection{First order approximation of the operator $\hat{O}(t,s,z)$}
\label{sectfirstorder}
We need to know the operator $\hat O(t,s,z)$ from (\ref{ooperator}) 
in order
to solve the non-Markovian QSD equation (\ref{nnm}) on a computer.
Notice that $\hat O$ enters (\ref{nnm}) under the memory
integral $\int_0^t\alpha(t,s)\hat O(t,s,\tilde z)\,ds$ only. 
Therefore, if the
correlation time of the environment is not too long, only $s$-values
in the vicinity of the upper integration range are relevant, and thus 
we consider the expansion of the operator
$\hat{O}(t,s,z)$ in Eq.(\ref{nnm}) in powers of $(t-s)$,
\beqa
\hO(t,s,z)&=&\hO(s,s,z) +
\left.\frac{\partial\hO(t,s,z)}{\partial t}\right\arrowvert_{t=s}(t-s)
\nonumber\\
&&+\left.\half\frac{\partial^2\hO(t,s,z)}{\partial
t^2}\right\arrowvert_{t=s}(t-s)^2+ ...
\label{exp}
\eeqa

Substituting  (\ref{exp}) into (\ref{nnm}) or (\ref{nnm1}), we
get a hierarchy of approximate QSD equations by truncating the above
expansion.
The validity of the corresponding approximation depends on the
environment correlation time $\tau$ determined by the correlation
function
$\alpha(t,s)$. In fact, it turns out that $\hat O$ changes on
{\it system} time scales as a function of $(t-s)$ and thus, the
expansion (\ref{exp}) corresponds to a systematic expansion
of the non-Markovian QSD equation in powers of the
number $\omega\tau$
where $\omega$ is a typical `system' frequency and $\tau$ the
environmental correlation time. For example,
the zeroth-order term leads to the standard Markov QSD, when
$\tau\rightarrow 0$.
The first-order term is the most important correction to the
Markov dynamics. Therefore, in what follows we will work out the
approximation up to the first order in some detail.

By using the consistency condition (\ref{consi}), one can work out the
following expressions for the operator $\hat{O}(t,s,z)$ at time
point $t=s$, without knowing its explicit form (for details, see Appendix
 A):
\beqa
\hat{O}(s,s,z) &=& L\label{pur1}\\
\left.\frac{\partial\hat{O}(t,s,z)}{\partial
t}\right\arrowvert_{t=s}&=&-i[H,L]
-\int_0^s\alpha(s,u)du [L^\da,L]L
\label{pur2}
\eeqa
where $H$ is the Hamiltonian and $L$ is the Lindblad operator, as
specified in the previous section. Now, we are in a position to write 
out the non-Markovian QSD equation up to the first order. Indeed, 
the first two terms in the expansion (\ref{exp}), substituted 
repectively by (\ref{pur1}) and (\ref{pur2}),
yield $\bar{O}(t)$ in (\ref{nota}) in the following form:
\beq
\bar{O}(t) = g_0(t)L - g_1(t)i [H,L] - g_2(t)[L^\da,L]L
\label{ob}
\eeq
where
\beqa
g_0(t)&=&\intt\al(t,s)ds\label{g0}\label{o1}\\
g_1(t)&=& \intt\al(t,s)(t-s)ds\label{g1}\label{o2}\\
g_2(t)&=& \intt\int^s_0\al(t,s)\al(s,u)(t-s)duds\label{o3}
\label{g2}
\eeqa
Note that $g_0$ is of the order one, yet $g_1$ and $g_2$ are of the 
order of the environmental correlation time $\tau$.
Substituting (\ref{ob}) into (\ref{nnm1}), the first-order 
non-Markovian nonlinear QSD equation is obtained:
\beqa
\frac{d}{dt}\tilde\psi_t=&-&iH\tilde\psi_t +\Delta_t(L)\tilde\psi_t\tilde
z_t\nonumber\\
&-&g_0(t)\left(\Delta_t(L^\da) L - \LJ \Delta_t(L^\da)L\RJ_t
\right)\tilde\psi_t\nonumber\\
&+&
ig_1(t)\left(\Delta_t(L^\da)[H,L]-\LJ\Delta_t(L^\da)[H,L]\RJ_t\right)\tilde
\psi_t\nonumber\\
&+&  g_2(t)\left(\Delta_t(L^\da)[L^\da,L]L -\LJ
\Delta_t(L^\da)[L^\da,L]L\RJ_t \right)\tilde\psi_t
\label{nqsd}
\eeqa
where $\tilde z_t$ is the shifted noise, $\Delta_t(L)=L-\LJ L\RJ_t$, and
$\LJ L\RJ_t=\LJ\tilde\psi_t\vert L\vert\tilde\psi_t\RJ$ is the quantum
expectation value. 

The Hamiltonian $H$ defines a typical system frequency $\omega$,
the combination $L^\dagger L$ defines a typical system relaxation 
rate $\Gamma$. We thus see that the zeroth order term in (\ref{ob}) gives
rise to a term of the order $\Gamma$ (second line in eq.(\ref{nqsd}),
whereas the two first order terms
in (\ref{ob}) lead to corrections which are smaller by a factor
$\omega\tau$ or $\Gamma\tau$, respectively 
(third and fourth line in eq.(\ref{nqsd}), $\tau$ is again the 
environment correlation time). Therefore, we expect (\ref{nqsd}) to
be valid for non-Markovian situations where the environmental
correlation time may be finite but no larger than typical
system time scales. The Markov case emerges for $\tau\rightarrow 0$,
where the first order correction becomes negligible and only the
zeroth order term remains. Then (\ref{nqsd}) reduces to the 
standard Markov QSD equation for $t>0$.

We also see that non-Markovian properties are encoded
in the time dependent coefficients $g_i(t)$ which change on the very fast
environmental correlation time scale $\tau$. The absence of the noise $z$ 
in the first order expansion (\ref{ob}) is remarkable (higher order
expansions contain the noise). Note that the approximate
non-Markovian QSD equation (\ref{nqsd}) still preserves the norm of the
wavefunction. Eq. (\ref{nqsd}) is the main result of this section. 
As stated before, applicability of non-Markovian
QSD lies in the determination of the operator $\hat{O}(t,s,z)$.
As we have already pointed out, the difficulties in handling
non-Markovian unravelings is often the nonlocal noise $z$ appearing either
in the functional derivative (see (\ref{qsd})) or in the integrand operator
$\hat{O}(t,s,z)$ (see (\ref{qsdn}) and (\ref{nnm})). 
We see that the above approximate QSD equation greatly simplifies
non-Markovian
QSD equation (\ref{nnm1}).

In addition, Eq.(\ref{nqsd}) is explicitly written in terms of
the Hamiltonian of the system $H$, the Lindblad operator $L$ and their
various
commutators. All of these can be obtained automatically once the physical
model
is specified. The only work left is to calculate the coefficients $g_i(t)
(i=0,1,2)$ from the environment correlation function $\alpha(t,s)$.       

After working out the formal perturbative QSD equation, it is useful to
see the concrete form of the coefficients $g_i(t)$. For
simplicity, we assume here that the system is driven by
Ornstein-Uhlenbeck noise, characterized by the exponential correlation
function
\beq
\al(t,s)=\frac{\ga}{2}\e^{-\ga\vert t-s\vert}
\eeq
where $\ga^{-1}=\tau$ defines the finite environmental memory
or correlation time. Note that this corresponds to a Lorentzian spectrum.
In the limit $\ga\rightarrow\infty$, the Ornstein-Uhlenbeck
noise reduces to simple comples white noise:
\beq
\al(t,s)=\de(t-s)
\label{del}
\eeq
In the case of the Ornstein-Uhlenbeck process, the coefficients $g_i(t)$ 
can be easily obtained from Eq. (\ref{o1})-(\ref{o3}):
\beqa
g_0(t)&=& \half\left(1-\e^{-\ga t}\right)\label{coe1}\\
g_1(t)&=& \frac{1}{2\ga}\left(1-\e^{-\ga t}-\ga t\e^{-\ga
t}\right)\label{coe2}\\
g_2(t)&=& \frac{1}{4\ga}
\left(1-\e^{-\ga t}-\ga t\e^{-\ga t}-\half \ga^2t^2\e^{-\ga t}\right)
\label{coe3}
\eeqa

In the long-time limit $t\gg\tau$, we see that the coefficients of 
the non-Markovian QSD
(\ref{nqsd}) become
constant: $g_0=1/2, g_1=1/2\ga, g_2=1/4\ga$, which also confirms that
$g_0$ is of the order one whereas $g_1$ and $g_2$ are of the order of
the environmental correlation time $\tau=\ga^{-1}$.

In the Markov limit $\ga\rightarrow\infty$,
$g_0(t)\rightarrow 1/2$ and $g_1(t), g_2(t)\rightarrow 0$ for $t>0$
and the non-Markovian QSD
equation (\ref{nqsd}) reduces to the standard Markov QSD equation
\cite{GP}(Note here
we write it in the Stratonovich form \cite{GG,Gar1}):
\beq
\frac{d}{dt}\tilde\psi_t=-iH\tilde\psi_t +
\Delta_t(L)\tilde\psi_t\circ(z_t + \LJ L^\da\RJ_t)
-\half\Delta_t(L^\da L)\tilde\psi_t
\eeq
with $z_t$ the standard complex white noise, as expected.

Our formal perturbation approach can be carried out to any desired order of
approximation
(For the details of the second order expansion
and the coefficients, see Appendix B). It is important to note, however,
that the
higher order derivatives of $\hat{O}(t,s,z)$ at
$t=s$ may contain the noise $z$.

Since the linear non-Markovian QSD equation (\ref{qsd}) is often simpler to
derive the corresponding master equation (see Section IV), we
also give its first-order approximation:
\beq
\dot\psi_t=-iH\psi_t + L\psi_t z_t - g_0(t)L^\da L\psi_t +
ig_1(t)L^\da[H,L]\psi_t
+ g_2(t) L^\da[L^\da,L]L\psi_t
\label{qsd1}                                            
\eeq
where the coefficients $g_0(t), g_1(t), g_2(t)$ are
given by (\ref{o1}), (\ref{o2}) and (\ref{o3}).

\subsection{Functional expansion of $\hat{O}(t,s,z)$}
In this subsection, we consider another kind of perturbative expansion,
the functional expansion of the operator $\hat{O}(t,s,z)$ in terms of noise
$z_v$:
\beqa
\hO(t,s,z)&=&\hO_0(t,s) \nonumber\\
&+&\int_0^t\hO_1(t,s,v)z_v dv \nonumber \\
&+&\int_0^t\int_0^t\hO_2(t,s,v_1,v_2)z_{v_1}z_{v_2}dv_1dv_2 \nonumber \\
&+& ....  \nonumber \\
&+& \int_0^t...\int_0^t
\hO_n(t,s,v_1,...,v_n)z_{v_1}...z_{v_n}dv_1...dv_n\nonumber\\
&+& ...
\label{fexp}
\eeqa
where the operators $\hO_n(t,s,v_1,...,v_n)$ are independent of the noise
$z$ and are
symmetric in their $n$ last variables (e.g.
$\hO_2(t,s,v_1,v_2)=\hO_2(t,s,v_2,v_1)$).
The initial condition is $\hO(t,t,z)=L$.
The expansion (\ref{fexp}) takes into account the generally 
nonlocal dependence of
the operator $\hat{O}(t,s,z)$ on the noise $z$.

From the consistency condition (\ref{consi}) and the
QSD equation (\ref{qsd}), we get a hierarchy
of equations for the operators
$\hat{O}_n(t,s, v_1, ..., v_n)$ (see Appenix A). 
Of particular interest is the zeroth order term $\hat{O}_0(t,s)$, which
satisfies
the following equation(ignoring the first order term):
\beqa
\frac{\partial}{\partial t}\hO_0(t,s)&=&[-iH, \hO_0(t,s)]\nonumber \\
&-&[L^\da\bO_0(t),\hO_{0}(t,s)]
\label{order1}
\eeqa
For the approximation $\hat{O}(t,s,z)\approx \hat{O}_0(t,s)$, the
 approximate
non-Markovain QSD equation then takes form:
\beq
\frac{d}{dt}\ti\psi_t=-iH\ti\psi_t +\Delta_t(L)\ti\psi_t\ti z_t
-\Delta_t(L^\da)\bar{O}_0(t)\ti\psi_t +
\LJ\Delta_t(L^\da)\bar{O}_0(t)\RJ_t\tilde\psi_t
\label{funqsd}
\eeq 
The justification of this approximation is that whenever the open quantum
system deviates
slightly from the Markov dynamics, then the first term $\hO_0(t,s)$ of
the expansion (\ref{fexp}) plays the dominant role. This can be easily
seen from the fact that all
$\bar{O}_n(t,v_1,...,v_n)\equiv\intt\al(t,s)\hat{O}_n(t,s,v_1,...,v_n)ds,
n\geq 1$
 go to zero in the Markov limit: $\al(t,s)\rightarrow\de(t-s)$,
except the first term $\bar{O}_0(t)\equiv\intt\al(t,s)\hat{O}_0(t,s)ds$
 which
goes to $\half L$. Physically this can be understood as follows. In the
Markov case, the
quanta coupled from the system to the environment never come back to the
system. Whereas, in the non-Markovian case, the emitted quanta will re-couple
from the environment to the system.

Similarly, one can build the higher order approximations which usually
 contain
the noise $z$. One obtains then a series of approximate QSD equations.
The master equation corresponding to the zeroth order approximation
(\ref{funqsd})
is derived in the next section.

\section{Non-Markovian QSD versus non-Markovian master equation}
\label{sec: ge}
In this section, we discuss how to derive the master equation from
the non-Markovian QSD equation. Our motivations are as follows. First, 
the master equation approach has a long tradition and is fundamental 
in open quantum system dynamics, and the reduced density operator 
contains all mean values of the `system' that can be directly observed 
and measured. Second, although it is clear in principle that each 
perturbative scheme for non-Markovian QSD gives rise to a perturbative scheme
for the non-Markovian master equation, it is very difficult in practice to
carry out this program without a systematic way to derive the non-Markovian
master equation from its QSD counterpart.  The aim of this  section is
to show how to derive the quantum master equation directly from
the non-Markovian QSD. Based on this result, we establish explicitly the
relation
between the perturbative QSD equations and perturbative master equations.
                                            
\subsection{General master equation}
The starting point of the derivation of the general master equation is  the
unnormalized
projection operator $P_t$,
\beq
P_t =\vert\psi_t(z)\rangle\langle\psi_t(z)\vert
\eeq
Recall that the reduced density operator can be reproduced by taking the
statistical
means over the noise: $\rh_t =
M[P_t]=M\left[\vert\psi_t(z)\rangle\langle\psi_t(z)\vert\right]$.
Accordingly, the temporal evolution equation for $P_t$ can then be obtained
from (\ref{qsdn}):
\beqa
{d\over dt}P_t =&-&i[H,P_t] +LP_tz_t +P_tL^\da z_t^{\ast}\nonumber\\
                &-& L^\da\intt\al(t,s) \hat{O}(t,s,z)ds P_t -
                P_t\intt\al(t,s)^\ast\hat{O}(t,s,z)^\da ds L
\label{proj}
\eeqa               
The above equation is, of course, a stochastic differential equation with
time-dependent
coefficients. Accordingly, the master equation corresponding to Eq.
(\ref{qsd}) may be
obtained by taking statistical mean values of Eq. (\ref{proj}).

To this end, we note that for any complex Gaussian noise $z_t$, the
following relations hold
(see Appendix C):
\beqa
M[P_tz_t]&=&\intt dsM[z_tz_s^{\ast}]M\left[\delta P_t\over \delta
z_s^{\ast}\right]
\label{fr}\\
M[P_t z_t^\ast]&=&\intt dsM[z_t^{\ast}z_s]
M\left[\delta P_t\over \delta z_s\right]\label{sr}
\eeqa
From (\ref{fr}) and (\ref{sr}), the following identities are obtained
\beqa
M[LP_tz_t]&=& L\intt\al(t,s)^\ast M\left[P_t\hat{O}(t,s,z)^\da\right] ds
\label{ans1} \\
M[P_tL^\da z_t^\ast]&=&
 \intt\al(t,s)M\left[\hat{O}(t,s,z)P_t\right]dsL^\da
\label{ans2}
\eeqa
Here we used the following relations:
\beqa
M\left[{\de\over \de z_s}P_t\right]&=&
M[{\de\over\de z_s}\vert\psi_t\rangle\langle\psi_t\vert]=
M[\hat{O}(t,s,z)P_t]\label{rel1}\\
 M\left[{\de\over \de z_s^\ast}P_t\right]&=&
M[\vert\psi_t\rangle{\de\over\de z_s^\ast}\langle\psi_t\vert]
=M[P_t\hat{O}(t,s,z)^\da]\label{rel2}
\eeqa
and we take advantage of the definition of the $\hat O$-operator
(\ref{ooperator}).
The validity of the above two identities (\ref{rel1}) and (\ref{rel2}) is
ensured by the fact that the solution $\psi_t$ of Eq. (\ref{qsd}) is the
analytic function
of $z$ and is thus independent of $z^\ast$. Accordingly,
$\de\vert\psi_t\RJ/\de z_t^\ast=0,~
\de\LJ\psi_t\vert/\de z_t=0$.

Hence, using (\ref{ans1}) and (\ref{ans2}), the exact non-Markovian master
equation
corresponding to non-Markovian QSD (\ref{qsd}) can be obtained:
\beq
{d\over dt}\rh_t
 =-i[H,\rh_t]+\left[L,M\left[P_t\bar{O}(t,z)^\da\right]\right]
                -\left[L^\da,M\left[\bar{O}(t,z)P_t\right]\right]
\label{proj1}
\eeq  
Where as before $M[\cdots]$ stands for the ensemble average, and
$\bar{O}(t,z)$ is defined in (\ref{nota}).

Eq. (\ref{proj1}) is the exact equation on which our perturbation approach
is based.
As an evolution equation, the above master equation does not look very nice
since the last two terms appearing in the equation have not yet been
written in terms of $\rh$.
It seems quite difficult to write this equation into a closed
evolution equation in full generality, if not impossible. We shall see,
however, that in many interesting and physically relevant situations, 
a closed
form for this equation can be found (see below).
Notably, the use of the relations (\ref{fr}) and (\ref{sr}) can make a
tremendous
simplification in deriving the master equation of open quantum system from
its QSD counterpart. In fact, it enables us to find out an exact or an
approximate
non-Markovian master equation by directly using the techniques of
stochastic process.

The non-Markovian master equation (\ref{proj1}), by design, will always
preserve
the positivity, trace and hermiticity.

\subsection{Approximate master equations}
Since the master equation (\ref{proj1}) cannot, in general, be written in a
closed form, some kind of approximation has to be made 
to determine the operator
$\hat{O}(t,s,z)$.
The Markov approximation emerges for a vanishing environment
correlation time, $\al(t,s)=\delta(t-s)$. 
In this case, from (\ref{ob}) $\bar{O}(t,z)=\frac{1}{2}L$, 
and Eq. (\ref{proj1})
reduces to the Markov Lindblad master equation,
\beq
{d\over dt}\rh_t =-i[H,\rh_t]+ L \rh_tL^\da
-\half\{LL^\da,\rh_t\}
\label{markov}
\eeq
Another interesting case is when the dependence of  the
 operator $\bar{O}(t,z)$ on the noise $z_t$ is
 negligible, that is,
$\bar{O}(t,z)\approx \bar{O}_0(t)$. Recall from (\ref{fexp}) that this is
indeed
the case when the dynamics is not far from Markov or the driving noise
is very small.
Under this approximation, the master equation takes the following simple
 form:
\beq
{d\over dt}\rh_t = -i[H,\rh_t]+ [L, \rh_t\bar{O}_0(t)^\da] +
[\bar{O}_0(t)\rh_t, L^\da]
\label{mass}
\eeq 
The notation $\bar{O}_0(t)$ is same as before(see
 (\ref{fexp}),(\ref{nota})).
The master equation (\ref{mass}) will serve as a good approximation
to the exact non-Markovian master equation (\ref{proj1}) in many situations
of
interests. In particular, if the operator $\hat{O}(t,s,z)$ is independent
of noise $z_t$, then Eq. (\ref{mass}) becomes exact. Interestingly,
there are many physically relevant examples
that satisfy this condition \cite{DGS}. 

More importantly for this paper, this condition is always satisfied in 
the first-order perturbative
approximation (\ref{ob}) developed in section \ref{sectfirstorder}. 
Then the master equation (\ref{proj1}) takes the
following
form:
\beqa
\frac{d}{dt}\rh_t =&-&i[H,\rh_t] +(g_0(t)+g^\ast_0(t))L\rh_t L^\da
-g_0(t)L^\da L\rho_t-g_0^\ast(t)\rh_tL^\da L\nonumber \\
&+& ig_1(t)[L^\da,[H, L]\rh_t] -ig^\ast_1(t)[\rh_t[L^\da,H],L]\nonumber \\
&+& g_2(t)[L^\da,[L^\da,L]L\rh_t] +g^\ast_2(t)[\rh_t L^\da[L^\da, L],L]
\label{mas1}
\eeqa
This master equation is the main result of this subsection. It provides a
systematic
evolution for first-order non-Markovian systems. Hence it could be called
the ``post-Markov" master equation. As for the first-order 
QSD equation (\ref{nqsd}), the second and third line
are smaller by a factor $\omega\tau$ or $\Gamma\tau$ compared to
the first line (recall that $\omega$ is the typical `system' frequency 
determined
by $H$, $\Gamma$ is a typical `system' relaxation time scale determined
by $L^\dagger L$, and $\tau$ is the environmental correlation time).

Note that this ``post-Markov" equation in general remains non-Markovian
even when $g_1(t)=0,g_2(t)=0$, because of the $g_0(t)$ term. However,
for long time $g_0(t)$ tends to a constant.

Equations (\ref{nqsd}) and (\ref{mas1}) will be  applied to some
examples in Sections V. In addition, Section VI presents
a perturbation analysis of the quantum Brownian motion model.

Finally, it shoud be noted that we have not touched issues such as
mathematical conditions for the convergence of the expansions (\ref{exp})
and (\ref{fexp}). Also, we are not able to prove, in full generality,
that Eq. (\ref{mas1}) always yields a positive density operator.
However, in this paper, we shall illustrate
in several examples that the resulting approximate QSD and master
equations around the Markov limit are mathematically consistent.
We will come back to these issues in future discussions.

\section{Examples and Applications}
\label{sec: exa}
The perturbative approach developed in the previous sections allows to
apply first order non-Markovian QSD to any open quantum system once the
Hamiltonian
of the system $H$, the Lindblad operator $L$ and the environment 
correlation
function $\al(t,s)$ are specified. All of these are determined by the
physical
model itself, as illustrated in this section using some typical models.
For simplicity, we assume that the complex process $z_t$ entering the
non-Markovian QSD equation (\ref{nqsd}) has a Lorentzian
spectrum, i.e. is of the Ornstein-Uhlenbeck type with the correlation
function $\al(t,s)=\frac{\ga}{2}\e^{-\ga\vert t-s\vert}$, where
$\ga^{-1}=\tau$ is the environmental correlation time,  unless otherwise
stated.

\subsection{Dissipative model}

In this subsection, we consider a dissipative two-level model characterized
 by
\beq
H=\frac{\om}{2}\si_z,\> \>L=\la\si_-\;.
\eeq
Since this model can be solved exactly\cite{DGS}, we are able to compare
the perturbation approach with the exact non-Markovian QSD and master
equations. Note that the model defines two `system' time scales
through the parameters by $\omega$ (oscillation) and $\lambda^2$ (damping).
Here we assume that they are of the same order of magnitude.

The first-order non-Markovian QSD equation can be obtained from
 (\ref{nqsd}):
\beqa
\frac{d}{dt}\tilde\psi_t=&-&i\frac{\om}{2}\si_z\ti\psi_t +
\la(\si_--\langle \si_-\rangle_t)\ti\psi_t\ti z_t\nonumber\\
&-&\left(\la^2g_0(t)+i\la^2\om g_1(t)+\la^4 g_2(t)\right)\left(\si_+ \si_-
\right.\nonumber\\
&&-\left. \LJ\si_+\RJ_t \si_- -\LJ\si_+\si_- \RJ_t +
\LJ \si_+\RJ_t\LJ\si_-\RJ_t\right)\ti\psi_t
\label{nqsd1}
\eeqa
and the first-order non-Markovian master equation can be obtained from
(\ref{mas1}):
\beqa
\frac{d}{dt}\rh_t =&-&i{\om\over 2}[\si_z,\rh_t]
+\la^2g_0(t)\left(2\si_-\rh_t\si_+ -
\{\si_+\si_-,\rh_t\}\right)\nonumber\\
&-&i\la^2\om g_1(t)[\si_+\si_-, \rh_t] -\la^4g_2(t)\{\si_+\si_-, \rh_t\}+
2\la^4g_2(t)\si_-\rh_t\si_+
\label{ze1}
\eeqa
where $g_1(t)$ gives the time-dependent frequency shift. Thus the master
equation is of the Lindblad form with time-dependent coefficients. As seen
in the next subsection, this property can not be regarded as a generic
 feature
of a non-Markovian master equation. Note that the first-order
master equation (\ref{ze1}) respects the hermiticity, normalization and
positivity for any initial states and time scales.
We can easily identify the first order non-Markovian corrections terms
of the order
$\omega/\gamma$ and $\lambda^2/\gamma$
in equations (\ref{nqsd1}) and (\ref{ze1}). We expect 
these equations to be a good approximation for the exact solution as
long as terms of the order
$(\omega/\gamma)^2$ and $(\lambda^2/\gamma)^2$ are negligible.

In Fig. 1, the average of $<\vec\sigma>$ for $\om=\la= 1$ and  $\ga=10$ 
are plotted.
The results given by perturbation
QSD equation over 2000 realizations (solid curve) is in remarkable
agreement with
the exact master equation (dotted curve).    

To illustrate the limits of the Markov approximation Fig. 2 presents the
ensemble
average $\LJ\si_x\RJ$ for the first-order QSD (solid curve)
for the same parameters as Fig. 1 except for the memory time $\ga=1$, and
compares this with the Markov master equation (dotted curve) and
the exact master equation (dashed curve). Clearly, the ensemble
average of $\si_x$ over 1000 trajectories still gives a good approximation
to the exact master equation. The result is fully in accordance
with our expectation as for relatively long memory times the Markov
approximation
is no longer valid. It should be noted that,
in general, the accuracy of the first-order QSD is also limited to
relatively short memory times,
but not as severly as the Markov approximation. Then the higher order
approximations or
alternative expansion such as (\ref{fexp}) should be used.

\subsection{Two-level model}\label{secttwolevelmodel}
Let us consider a driven two-level atomic system interacting with a
dissipative environment. The Hamiltonian of the system, $H$, and
the Lindblad operator, $L$, which represents the
influence of the environment are given by
\beq
H=\frac{\om}{2}\si_x,\> \>L=\la\si_z
\eeq
respectively, where the parameter $\la$ is a coupling constant. For this
model,
it can be shown that the expansion (\ref{fexp}) will not terminate at any
finite order. The application of the perturbative approach is thus
 necessary.
The first-order non-Markovian QSD equation can be readily obtained
from (\ref{nqsd}):
\beqa
\frac{d}{dt}\tilde\psi_t=&-&i\frac{\om}{2}\si_x\ti\psi_t +
\la(\si_z-\langle \si_z\rangle_t)\ti\psi_t\ti z_t\nonumber\\
&+&\la^2g_0(t)\left(\LJ\si_z\RJ_t \si_z -
\LJ \si_z\RJ_t^2\right)\ti\psi_t\nonumber\\
&-& \om \la^2g_1(t)\left(i\si_x +\LJ\si_z\RJ_t\si_y -i\LJ\si_x\RJ_t -
\LJ\si_z\RJ_t\LJ\si_y\RJ_t\right)\tilde\psi_t
\label{nqsd2}
\eeqa
where the coefficients $g_0(t),g_1(t)$ are given by (\ref{coe1}) and
(\ref{coe2}),
respectively. The first two lines in the above equation are expected from
the Markov QSD picture. The third line represents the non-Markovian
correction and is smaller by a factor $\omega\tau=\omega/\gamma$.

Similarly, the  first-order non-Markovian master equation can be obtained
directly
from (\ref{mas1}):
\beqa
\frac{d}{dt}\rh_t=&-&i{\om\over 2}[\si_x,\rh_t]
 +2\la^2g_0(t)\si_z\rh_t\si_z-
2\la^2g_0(t)\rh_t\nonumber \\
&-&i\la^2\om g_1(t)[\si_x,\rh_t] -\la^2\om g_1(t)\si_z\rh_t\si_y
-\la^2\om g_1(t)\si_y\rh_t\si_z
\label{zm}
\eeqa
There are some new features about the master equation (\ref{zm}).
First, it is obviously not in the Lindblad form due to the presence
of the cross term $\si_z\rh\si_y$ and its conjugate. Second, the
master equation derived in this way naturally preserves the hermiticity,
trace and positivity. The preservation of trace and hermiticity is obvious.
It is known that positivity
of any two dimensional density matrix is equivalent to the condition
$\vert\vert<\vec\sigma>\vert\vert\leq 1$, where 
$<\vec\sigma>={\rm Tr}(\vec\sigma\rh)$ is the
Bloch vector \cite{NG, KO}. In Fig. 3 we have plotted the norms of the
Bloch vector using the time-dependent master equation (\ref{zm}) (solid
curve) and
the long-time limit master equation(LME) (dotted curve), in which the
coefficients of the master equation become constant: $g_0(t)=1/2,
g_1(t)=1/2\gamma$.
Clearly, LME loses positivity for some initial states at short time scales,
whereas, the time-dependent master equation (\ref{zm}) preserves postivity
 at
all times. Note that this simple model is the 2-level analog of the
Caldeira-Leggett master equation studied in Section VI below.

We also solved numerically the first-order QSD equation (\ref{nqsd2}).
The average of $\vec{\si}$ obtained through many realizations of
(\ref{nqsd2}) (solid line) and through the first-order master
equation (\ref{zm}) (dotted curve) are plotted in Fig. 4. Taking the
ensemble
mean over 500 realizations we see from Fig. 4 that the first order QSD
equation is in good agreement with the first-order master equation, for the
short memory time($\ga=10$).

\section{Quantum Brownian Motion: Perturbative analysis}\label{QBM}
The transition from non-Markovian to Markov processes is an
outstanding problem. It is debated how to take the correct Markov limit
for a non-Markovian process. Certain approximations of the exact
dynamics can lead to master equations with bad properties such as
non-positivity. A notorious example is the Caldeira-Leggett master
equation \cite{CaLe,UZ,HPZ} which may violate positivity of the density
operator at short time scale \cite{Am,Pec,Haake,DD}. Consequently, it is
impossible to simulate friction \`a la Caldeira-Leggett with
stochastic Schr\"odinger equations. The aim of this section is to apply
the time-dependent perturbation approach for master equation developed
in the previous sections to the Quantum
Brownian Motion (QBM) model \cite{CaLe,UZ,HPZ,HY}.
In particular, we shall show that our first-order non-Markovian
master equation recovers the Caldeira-Leggett master equation in
the Fokker-Planck and long-time ($t\gg\tau$) limit.
The Hamiltonian of the system and the Lindblad operator are as follows:
\beq
 H=\frac{1}{2}p^2+ V(q), \> \>  L= q
\label{qbm}
\eeq
where we choose a unit mass particle moving in a general potential
$V(q)$. For the sake of simplicity, we consider the case of
the Ohmic heat bath, $I(\om)\sim \om$. The bath
correlation function is then given by
 \beq
 \al(t,s)={\eta\over\pi}\int^\Lambda_0d\om\om\left(\coth({\om\over 2kT})
 \cos[\om(t-s)]-i\sin[\om(t-s)]\right)
 \label{ohm}
 \eeq
where $\Lambda$ is the cut-off frequency of the bath which
characterizes the correlation time $\tau=\Lambda^{-1}$
and $\eta$ is the friction coefficient.

From (\ref{ob}), we get
\beq
\bar O(t)=g_0(t)q-g_1(t)p
\eeq
where the coefficients $g_0(t), g_1(t)$ are defined as before
(\ref{o1}),(\ref{o2}).

The zeroth-order master equation can be obtained from (\ref{mas1})
by setting $g_1(t)=g_2(t)=0$:
\beq
 {d\over dt}\rh_t = -i[H,\rh_t] -g_{0R}(t)[q,[q,\rh_t]]
 -ig_{0I}(t)[q^2,\rh_t].
\label{ze}
 \eeq
This master equation preserves positivity for all times, regardless
of the initial states, as it is of 
the standard Lindblad form with time dependent coefficients.
(This is of course not a generic feature
for non-Markovian master equations).
However, Eq. (\ref{ze}) 
does not take the energy dissipation into account.
More relevant is therefore the first order approximation.
The master equation in this case can be obtained from (\ref{mas1})
\beqa
{d\over dt}\rh_t =&-&i[H,\rh_t] - g_{0R}(t)[q,[q,\rh_t]]
-ig_{0I}(t)[q^2,\rh_t]\nonumber\\
 &+&g_{1R}(t)[q,[p,\rh_t]]+ig_{1I}(t)[q,\{p,\rh_t\}]
\label{fi}
\eeqa
where the coefficients $g_{iR}(t), g_{iI}(t)~ (i=0,1)$ are the real and
imaginary parts of $g_{i}(t),(i=0,1)$, respectively.
The coefficient $g_{0R}(t)$ induces
diffusion and the decoherence in positon $q$ while $g_{0I}(t)$ gives
rise to a time-dependent frequency shift. The coefficient $g_{1R}(t)$ is
responsible for further diffusion, and the last coefficient $g_{1I}(t)$
gives the friction. All of these time-dependent coefficients vanish at
$t=0$ due to the assumption that initially the state of bath+system is
factorable. In the special case when $V(q)$ is a quadratic potential, it is
reassuring that our non-Markovian master equations (\ref{ze}) and (\ref{fi})
coincide with the zeroth and, respectively, first order expansions of
the exact Hu-Paz-Zhang master equation \cite{HPZ}.

In the Ohmic case (\ref{ohm}), there exists a special high
temperature limit (Fokker-Planck limit) which results in a Markov master
equation. We take the high temperature limit
in such a way that $kT\gg \Lambda$. For times
$t\gg \tau=\Lambda^{-1}$
the time dependent coefficients in (\ref{fi}) approach constant values
and we get
%
the Markov `Caldeira-Leggett' master equation for Brownian motion:
\beq
 {d\over dt}\rh_t = -i[H^\prime, \rh_t] -i\frac{\eta}{2}[q, \{p,\rh_t \}] 
-
\eta
kT[q, [q,\rh_t]]  
\label{cale}
\eeq 
where $H^\prime$ is the cutoff-dependent renormalized Hamiltonian.
This is a Markov master equation with constant coefficients. It does not
belong to the Lindblad class and it may violate the positivity of 
the density operator. We mention
casually that a next order high temperature expansion improves this
situation and replaces the Caldeira-Leggett equation
(\ref{cale}) by a proper Markov Lindblad equation \cite{DD}. This $1/T$
asymptotic expansion has nothing to do with the perturbative approach in
our present work.  
Note that the Fokker-Planck limit of the zeroth-order non-Markovian equation
(\ref{ze}) does not contain the dissipative (friction) term on the r.h.s.
so it is a Lindblad master equation. 

It is instructive to look at the non-positivity of the Caldeira-Leggett
master equation (\ref{cale}) from the QSD point of view. It is clear 
from the derivation that the QSD master equation (\ref{fi}) differs 
from the
standard Caldeira-Legget master equation (\ref{cale}) for short times
of the order of the environmental correlation time.
During this short 
time, an arbitrary initial condition, which might lead to positivity
violation when propagated with the non-Lindblad master equation 
(\ref{cale}), evolves towards an effective, modified
`initial' density operator for the long time master equation (\ref{cale})
\cite{Haake}.

Our QSD master equation (\ref{fi}) is also a non-Lindblad
equation but with time-dependent coefficients. 
As in the case
of the spin model in section \ref{secttwolevelmodel}, their time dependence
can assure the preservation of the state's positivity.
In the master equation (\ref{fi}), the coefficient $g_{1I}(t)$ of the
dissipative term is zero at $t=0$ and its time derivative vanishes, too. 
The diffusion coefficient $g_{0R}(t)$ also vanishes
but its initial derivative is positive. Thus the initial phase of the
evolution is dominated by diffusion. This mechanism
may, as is well known in the exact model of Ref.\cite{HPZ}, guarantee
the positivity of the density matrix at short times as well as at later
 times
when the dissipation enters. In contrast, in the Caldeira-Leggett
master equation (\ref{cale}) the constant dissipative term will immediately
violate the positivity of a distinguished class of initial density matrices.

In summary, we have presented the zeroth-order master equation (\ref{ze})
and the first-order non-Markovian master equations (\ref{fi}) based on
QBM model. After an initial `slip' time, of the order of the
environmental correlation time, we recover the standard QBM master
equation. We note that both decoherent histories and environment-induced
decoherence
are discussed using the QBM model, but mainly in the
Markov regimes \cite{DGHP,DoHa,GeHa,PaZu}. It would be interesting to
study these approaches with non-Markovian master equation like 
(\ref{fi}). We shall discuss these topics elsewhere.

\section{Conclusions}
Non-Markovian QSD offers a brand new avenue to explore non-Markovian
dynamics of open quantum systems. Such situations appear in a variety of
practical problems, like e.g. materials with photonic bandgaps 
or output coupling from a Bose-Einstein condensate. In this
article we present a systematic perturbation approach to non-Markovian
QSD.

Our perturbation approach makes non-Markovian QSD more amenable to
computer simulations. In particular, a detailed analysis of
first-order ``post-Markov" QSD equations and the corresponding
 ``post-Markov"
master equations are presented in Sections III and IV. It is noteworth that
these equations depend only on the system Hamiltonian, the Lindblad
operator and the environment correlation function. The equations can thus
be read
off directly from the total system+environment Hamiltonian. We have
illustrated the perturbation approach with some typical examples.

In the Markov regime, it is well-known that each Lindblad master
 equation
can be unraveled by either continuous or jump trajectories which decompose
the density matrix into pure states at all times. The reverse is also
true, each stochastic unraveling uniquely yields a positive density matrix.
In the present paper, we have shown that this correspondence is even more
fruitful
in the non-Markovian regime. We show explicitly how the non-Markovian QSD
equation gives rise to the corresponding non-Markovian master equation.
As the most important application, we have shown that each perturbative 
QSD equation
naturally gives rise to a perturbative master equation. We have shown
numerically that the resulting master equation naturally respects
the properties of hermiticity, normalization and, more importantly,
positivity.

Admittedly, many issues remain to be solved in this subject. In this 
paper we have exclusively discussed the first-order ``post-Markov''
perturbation theory for QSD without
touching the perturbative QSD based on the functional
expansion (\ref{fexp}). It is important to note that these
two expansions (\ref{exp}) and (\ref{fexp}) are of rather different
physical meaning. 
The former expansion, on which we concentrate in this paper is
an expansion in the environmental correlation time.
whereas  the latter is the expansion for the 'small noise'.
Clearly, the comparison of these two expansions will be interesting.
Another important project in the next step is to apply the non-Markovian
QSD to some realistic physical problems such as
non-Markovian atom-field interaction and in particular, the superradiance
near a photonic band gap, in which the non-Markovian interaction is
essential (e.g., see \cite{NJ}). Also, it is known in the
Markov regime \cite{SBP1,SBP2} that localization
of quantum trajectories - typically in phase space - 
is of great significance in accelerating the numerical simulations.
Therefore, investigations into
localization
in non-Markovian QSD would be useful in both theoretical
and practical respects.

\section*{Acknowledgements}
We particularly thank Ian Percival for reading the manuscript and for 
many useful suggestions. We also thank Fritz Haake, Jonathan Halliwell,
Bei-Lok Hu, Karsten Jakobsen and Klaus M{\o}lmer for stimulating
discussions and useful communications.
TY and NG would like to thank for support from the
Swiss National
Science Foundation and the European TMR Network ``The physics of quantum
information" (via the Swiss OFES).
LD is partly supported by the
 Hungarian
Scientific Research Fund through Grant No. T16047.
 WTS thanks the Deutsche
Forschungsgemeinschaft for support through the Sonderforschungsbereich 237
``Unordnung und gro{\ss}e  Fluktuationen''.

\begin{appendix}
\section{Perturbation expansion of the operator $\hat{O}(t,s,z)$}
Let us consider the following expansion of the operator $\hat{O}(t,s,z)$
\beqa
\hO(t,s,z)&=&\hO_0(t,s) \nonumber\\
&+&\int_0^t\hO_1(t,s,v)z_v dv \nonumber \\
&+&\int_0^t\int_0^t\hO_2(t,s,v_1,v_2)z_{v_1}z_{v_2}dv_1dv_2 \nonumber \\
&+& ....  \nonumber \\
&+& \int_0^t...\int_0^t
\hO_n(t,s,v_1,...,v_n)z_{v_1}...z_{v_n}dv_1...dv_n\nonumber\\
&+& ...
\label{exp1}
\eeqa
where the operators $\hO_n(t,s,v_1,...,v_n)$ are independent of the noise
$z$ and are
symmetric in their n last variables (eg
$\hO_2(t,s,v_1,v_2)=\hO_2(t,s,v_2,v_1)$).
The initial condition is $\hO(t,t,z)=L$.

Accordingly, we get
\beqa
\frac{d}{dt}\frac{\delta\psi_t}{\delta z_s}&=&
 \pt\hO_0(t,s)\psi_t\nonumber \\
&+&\left(\hO_1(t,s,t)z_t + \int_0^t\pt\hO_1(t,s,v)z_vdv \right)\psi_t
\nonumber \\
&+&\left(2\int_0^t\hO(t,s,t,v_2)z_tz_{v_2}dv_2 +
\int_0^t\int_0^t\pt\hO_2(t,s,v_1,v_2)z_{v_1}z_{v_2}dv_1dv_2 \right)\psi_t
\nonumber \\
&+& ....  \nonumber \\
&+&\left(n\int_0^t...\int_0^t\hO(t,s,t,v_2,...,v_n)z_tz_{v_2}...z_{v_n}dv_2.
..dv_n \right.\nonumber\\
&+& \left. \int_0^t...\int_0^t
\pt\hO_n(t,s,v_1,...,v_n)z_{v_1}...z_{v_n}dv_1...dv_n \right)\psi_t + ...
\nonumber \\
&+&\hO(t,s,z)\frac{d}{dt}\psi_t
\eeqa
and
\beqa
\frac{\delta}{\delta
z_s}\frac{d}{dt}\psi_t&=&\left(-iH+Lz_t\right)\frac{\delta\psi_t}{\delta
z_s}\nonumber\\
&-&L^\da\bO(t,z)\frac{\delta\psi_t}{\delta z_s} \nonumber \\
&-&L^\da\left(\bO_1(t,s)+2\int_0^t\bO_2(t,s,v_2)z_{v_2}dv_2+...\right.
\nonumber\\
&+&\left. n\int_0^t...\int_0^t
\bO_n(t,s,v_2,...,v_n)z_{v_2}...z_{v_n}dv_2...dv_n
\right)\psi_t + ...
\eeqa
where
 $\bO_n(t,v_1,...,v_n)\equiv\int_0^t\alpha(t,s)\hO_n(t,s,v_1,...,v_n)ds$.

Consequently, from the consistency condition
\beq
\frac{d}{dt}\frac{\delta}{\delta z_s}\psi_t=\frac{\delta}{\delta
z_s}\frac{d}{dt}\psi_t
\eeq
one obtains the following hierarchy of equations:
\beqa
\frac{\partial}{\partial t}\hO_n(t,s,v_1,...,v_n)&=&[-iH,
\hO_n(t,s,v_1,...,v_n)]\nonumber \\
&-&\frac{1}{n!}\sum_{P_n\in S_n}\sum_{k=0}^n
[L^\da\bO_k(t,v_{P_n(1)},...,v_{P_n(k)}),
\hO_{n-k}(t,s,v_{P_n(k+1)},...,v_{P_n(n)})] \nonumber \\
&-& (n+1)L^\da\bO_{n+1}(t,s,v_1,...,v_n)
\label{per1}
\eeqa
with initial conditions:
\beq
\hO_0(t,t)=L
\eeq
\beq                                  
\hO_n(t,t,v_1,...,v_n)=0 ~~for~ all~ n\ge1
\eeq
\beq
n\hO_n(t,s,t,v_2,...,v_n)=[L, \hO_{n-1}(t,s,v_2,...,v_n)]
\eeq
Where $S_n$ is the permutation group and $P_n$ is the permuation operators
acting on
the indices $v_1,v_2, ...,v_n$.

Of particular interest is $n=0$, we get
\beqa
\frac{\partial}{\partial t}\hO_0(t,s)&=&[-iH, \hO_0(t,s)]\nonumber \\
&-&[L^\da\bO_0(t),\hO_{0}(t,s)] \nonumber \\
&-& L^\da\bO_1(t,s)
\label{deriv}
\eeqa
From this the derivative of $\hat{O}_0$ can be easily worked out.

From simplicity, we assume here exponentially decaying
correlations:
\beq
\alpha(t,s)=\frac{\gamma}{2}e^{-\gamma|t-s|}
\eeq   
The evolution equations for the $\bO_n(t,v_1,...,v_n)$ read:
\beqa
\frac{\partial}{\partial t}\bO_n(t,v_1,...,v_n)
&=&\frac{\gamma}{2}{\hO}_n(t,t,v_1,...,v_n) -
\gamma\bO_n(t,v_1,...,v_n)\nonumber \\
&+&[-iH, \bO_n(t,v_1,...,v_n)] \nonumber \\
&-&\frac{1}{n!}\sum_{P_n\in S_n}\sum_{k=0}^n
[L^\da\bO_k(t,v_{P_n(1)},...,v_{P_n(k)}),
\bO_{n-k}(t,v_{P_n(k+1)},...,v_{P_n(n)})] \nonumber \\
&-& (n+1)L^\da\int_0^t\alpha(t,s)\bO_{n+1}(t,s,v_1,...,v_n)ds
\label{per2}
\eeqa
where $n\bO_n(t,t,v_2,...,v_n)=[L, \bO_{n-1}(t,v_2,...,v_n)]$ for $n\ge 1$
 and
$\bO_n(0,v_1,...,v_n)=0$ for all $n$.

The Eqs. (\ref{per1}) and (\ref{per2}) are very useful in the determination
of the operator
 $\hat{O}(t,s,z)$.

\section{Second order QSD equation}
In this appendix, we present the second order non-Markovian QSD
equation.

By using the functional expansion of $\hat{O}(t,s,z)$ and the consistency
condition
(See Appendix A), we can work out expansion of
the operator $\hat{O}(t,s, z)$ at point $s$ to any desired order. In what
follows,
for simplicity, we only give the second order expansion of the operator
$\hat{O}_0(t,s)$,
which contains no nonlocal noise $z$.
\beqa
\hat{O}_0(s,s) &=& L\\
\partial_1\hat{O}_0(s,s)&=&-i[H,L]-g_0(s)[L^\da,L]L\\
\partial_1^2\hat{O}_0(s,s)&=&-[H,[H,L]]+ig_0(s)[H,[L^\da,L]L]-\al(s,s)
[L^\da,L]L\nonumber\\
&&+ ig_0(s)[[L^\da[H,L],L]+g_0^2(s)[L^\da[L^\da,L]L,L]\nonumber\\
&&+ ig_0(s)[L^\da L,[H,L]] +g_0^2(s)[L^\da L,[L^\da,L]L]
\label{plau}
\eeqa
Note that all derivatives above are kinds of approximations, in particular,
the second
order derivative might contain more terms. Taking the first three terms of
the expansion (\ref{exp}), and making the
approximation $\hat{O}(t,s,z)\approx\hat{O}_0(t,s)$,
one obtains
\beqa
\bar{O}_0(t) &=& g_0(t)L-ig_1(t)[H,L]-g_2(t)[L^\da,L]L\nonumber\\
&&-g_3(t)[H,[H,L]] -g_4(t)[L^\da,L]L\nonumber\\
&&+ig_5(t)\left([H,[L^\da,L]L]+[L^\da[H,L],L] +[L^\da
L,[H,L]]\right)\nonumber\\
              && + g_6(t)\left([L^\da[L^\da,L]L,L]+[L^\da
L,[L^\da,L]L]\right)
\label{aa}
\eeqa
where the coefficients are as follows:
\beqa
g_0(t)&=&\intt\al(t,s)ds\\
g_1(t)&=& \intt \al(t,s)(t-s)ds \\
g_2(t)&=&\intt\int_0^s \al(t,s)\al(s-u)(t-s)dsdu\\
g_3(t)&=& \half\intt\al(t,s)(t-s)^2ds  \\
g_4(t)&=&\half\intt\al(t,s)\al(s,s)(t-s)^2ds\\
g_5(t)&=&\half\intt\int^s_0\al(t,s)\al(s-u)(t-s)^2duds\\
g_6(t)&=&\intt\int^s_0\int^u_0\al(t,s)\al(s,u)\al(s,v)(t-s)^2dvduds
\eeqa
Then the second order QSD equations  can be obtained by
substituting (\ref{aa}) into (\ref{nnm1}).  Notice that,
in principle, we could obtain any order approximate QSD
equations by directly using the consistency
condiiton and the functional expansion of
$\hat{O}(t,s,z)$.

\section{Derivation of the relations (\ref{fr}) and (\ref{sr})}
In this appendix, we shall prove the relations (\ref{fr}) and (\ref{sr}).
We take (\ref{fr}) for instance:
\beq
M[P_tz_t]=\int dsM[z_tz_s^\ast]M\left[\frac{\de P_t}{\de z_s^\ast}\right]
\label{re}
\eeq

Suppose the complex Gaussian measure takes the form:
\beq
 P(z)d\mu = N\exp\left[-\int d\si\int d\tau z^\ast_\si z_\tau
\beta(\si,\tau)\right]d\mu
\eeq
here $N$ is the normalization constant, and $\beta(\si,\tau)$ is a kernel
reciprocal to
the correlation function $\al(\la,\tau)$, which is defined by
\beq
M[z_t^\ast z_s]=\al(t,s)
\eeq
Note that the correlation function $\al(t,s)$ satisfies
$\al(t,s)=\al(s,t)^\ast$. We
then have following relation:
\beq
\int \al(t,\tau)^\ast\beta(\tau,s)d\tau=\de(t-s)
\label{dre}
\eeq
Now, we consider the right hand side of (\ref{re}):
\beqa
\int ds M[z_tz_s^\ast]M\left[\frac{\de P_t}{\de z_s^\ast}\right]
&=&N\int d\mu \int ds\al(t,s)^\ast\frac{\de P_t}{\de z_s^\ast}P(z)
 \nonumber\\
&=& -N\int d\mu\int ds\al(t,s)^\ast P_t\frac{\de}{\de z_s^\ast}P(z)
\label{dre1}
\eeqa
Here, integration by parts has been used from the first line to the second
line. Note that
\beq
\frac{\de}{\de z_s^\ast }P(z)=-\int d\tau z_\tau\beta(s,\tau)P(z)
\label{par}
\eeq
Inserting (\ref{par}) into (\ref{dre1}), changing the integration order
$\int ds$
and $\int d\tau$, and using the relation (\ref{dre}), we obtain (\ref{re}).
The relation (\ref{fr}) can be obtained by taking the hermitian conjugate
of (\ref{re}).

\end{appendix}

\newpage
\section*{Captions of Figures}
Fig. 1. Ensemble average of the Bloch vector $<\vec\si>$ over 2000
trajectories of
first-order QSD (solid curve), and by exact master equation (dotted curve)
with
Hamiltonian $H=(\om/2)\si_z$, $L=\la \si_-$ and
$\al(t,s)=\frac{\ga}{2}\e^{-\ga|t-s|}$. Here we
choose $\om=\la=1$ and $\ga=10$. The initial state is chosen as
$|\psi_0\RJ=|-\RJ +i|+\RJ$. 
\section*{}
Fig. 2. Ensmble average of $<\si_x>$ over 1000 realizations by using the
first-order
non-Markovian QSD
(solid curve) for the same model as Fig. 3. Here, $\gamma=1, \om=\la=1$
and $|\psi_0\RJ=|-\RJ+3|+\RJ$.
The dashed curve is the exact mster equation for the same choice of
parameters, and
the dotted curve is the master equation in Markov limit .

\section*{}
Fig. 3. Illustrtion of the norm of Bloch vector of two-level system with
$H=(\om/2)\si_x, L=\la\si_z$ and expontially decaying correlation
 function
$\al(t,s)=\frac{\ga}{2}\e^{-\ga|t-s|}$. The initial state is chosen as
the excited state $|\psi_0\RJ=|+\RJ$. The parameter are chosen as
$\om=\la=1, \ga=1/2$.
The solid curve represnts the norm of the Bloch vector by master equation.
And the dotted line represnts the norm of Bloch vector from long-time limit
master
equation (LME). We see $\vert\!\vert <\vec\si>\vert\!\vert>1$ for the Bloch
vector by
LME for the chosen initial state at short times. Accordingly, LME loses the
positivity
at short times for some initial states.

\section*{}
Fig. 4. Ensemble average of $<\vec{\si}>$ over 500 realizations (solid
 curve)
for the same model as Fig. 3. Here we choose $\om=\la=1, \ga=10$ and the
initial
state $|\psi_0\RJ=|-\RJ + \sqrt{3}|+\RJ$. The dotted curve is an average by
the first-order
master equation for the same choice of parameters.


\newpage
\begin{thebibliography}{99}
\bibitem{DGS}L, Di\'{o}si, N.\ Gisin and W.\ T. Strunz, Phys. Rev.{\bf A
58}, 1699 (1998);
       see also W. Strunz {\it Open System Dynamics with Non-Markovian
Quantum Trajectories},
           quant-ph 9803079.
\bibitem{BLM}S.\ Bay, P.\ Lambropoulos and K.\ M{\o}lmer, Phys. Rev. Lett.
{\bf 76},
161 (1996).
\bibitem{Ho}J.\ J. Hope, Phys. Rev. {\bf A 55}, R2531 (1997).
\bibitem{JQ}S.\ John and T.\ Quang, Phy. Rev. Lett. {\bf 74}, 3419(1994).
\bibitem{NJ}N.\ Vats and S.\ John, Phys. Rev. {\bf A 58}, 4168 (1998) and
reference therein.
\bibitem{MHS}G.\ M.\ Moy, J.\ J.\ Hope and C.\ M.\ Savage, quant-ph/9801046
(unpublished).
\bibitem{DM}J.\ Dalibard, Y.  Castin, and K.\ M{\o}lmer, Phys. Rev. Lett.
{\bf 68}
580 (1992).
\bibitem{C}H.\ Carmichael, {\it An Open Systems Approach to Quantum Optics}
(Springer-Verlag, Berlin, 1993).
\bibitem{PK}M.\ B.\ Plenio and P.\ Knight, Rev. Mod. Phys. {\bf 70}, 101
(1998)
and reference therein.
\bibitem{GP}N.\ Gisin and I.\ Percival, J. Phys. {\bf A 25}, 5677 (1992);
{\bf 26},
2233 (1993); {\bf 26}, 2243 (1993).
\bibitem{Perci}I.\ Percival, {\it Quantum State Diffusion} (Cambridge
University Press,
England, 1998).
\bibitem{WiMi}H.\ M.\ Wiseman and G.\ J.\ Milburn, Phys. Rev. {\bf A47},
1652 (1993).
\bibitem{CZA}C.\ Cohen-Tannoudji, B.\ Zambou, and E.\ Arimondo, J. Opt.
Soc. Am. {\bf B 10}, 2107 (1993).
\bibitem{P1}I.\ Percival, J. Phys. {\bf A 27}, 1003 (1994).
\bibitem{DGHP}L.\ Di\'{o}si, N.\ Gisin, J.\ J.\ Halliwell, and I.\
Percival, Phys. Rev.
Lett. {\bf 74}, 203 (1995).
\bibitem{BYu}T.\ A. Brun, Phys. Rev. Lett. {\bf 78}, 1833 (1997);
 T.\ Yu, Physica {\bf A 248}, 393 (1998).
\bibitem{Pe}P.Pearle, in {\it Perspectives on Quantum Reality}, edited by
Clifton
{\it et al.} (Kluwer Academic, Dordrecht, 1996)
 \bibitem{D}L.\ Di\'osi, Quantum Semiclasic. Opt. {\bf 8}, 309 (1996).
\bibitem{S}W.\ T.\ Strunz, Phys. Lett. {\bf A 224}, 25 (1996).
\bibitem{DS} L.\ Di\'{o}si and W.\ T.\ Strunz, Phys. Lett. {\bf A 235}, 569
(1997).
\bibitem{Im}A. Imamoglu, Phys. Rev. {\bf A 50}, 3650 (1994).
\bibitem{Ga}B. Garraway, Phys. Rev. {\bf A 55}, 2290 (1997).
\bibitem{Ga1}B. Garraway, Phys. Rev. {\bf A 55}, 4636 (1997).
\bibitem{BaLaMo}B.\ M.\ Bay, P.\ Lambropoulos, and K.\ M{\o}lmer, Phys.
Rev.Lett.
{\bf 79} 2654 (1997).
\bibitem{JCW}M.\ W. Jack, M.\ J.\ Collet, and D.\ F.\ Walls,
 quant-ph/9807028
(unpublished).
 \bibitem{CaLe}  A.\ O.\ Caldeira and A.\ J.\ Leggett, Physica {\bf A 121},
587 (1983).
\bibitem{GG}D.\ Gatarek and N.\ Gisin, J. Math. Phys. {\bf 32}, 2152 (1991).
\bibitem{Gar1}C.\ W.\ Gardiner, {\it Handbook of Stochastic Methods},
(Springer-Verlag,
Berlin, 1985).                           
\bibitem{NG}N.\ Gisin, Helv. Phys. Acta {\bf 63}, 930 (1990).
\bibitem{KO}A.\ Kossakowski, Bull. Adac. Polon. Sci., Ser. Sci. Math.
Astronom. Phys. {\bf 21}, 649 (1973).
\bibitem{UZ}W.\ G. Unruh and W.\ Zurek, Phys. Rev. {\bf D 40}, 1071 (1989).
\bibitem{HPZ}  B. L. Hu, H. P. Paz and Y. Zhang, Phys. Rev. {\bf D45}, 2843
(1992).
\bibitem{Am}V.\ Ambegaokar, Ber. Bunsenges. Phys. Chem. {\bf 95}, 400
 (1991).
\bibitem{Pec}P. Pechukas, in {\it Large-Scale Molecular Systems}, eds.:
W.Gans {\it et al.} (Plenum Press, New York, 1991).
\bibitem{Haake}S.\ Gnutzmann and F.\ Haake, Z. Phys. {\bf B 101}, 263 
 (1996).
\bibitem{DD}L.\ Di\'osi, Physica {\bf A 199}, 517 (1993);L.\ Di\'osi,
Europhys. Lett. {\bf 22}, 1(1993).
\bibitem{HY} J.\ P.\ Paz, in {\it The Physical Origin of Time
Asymmetry}, edited by J.\ J.\ Halliwell, J.\ Perez-Mercader,
and W.\ Zurek (Cambridge University Press, England, 1994); J.\ Halliwell
and T.\ Yu, Phys. Rev. {\bf D 53}, 2012 (1996).
\bibitem{DoHa}H.\ Dowker and J.\ Halliwell, Phy. Rev. {\bf D 46}, 1580
 (1992).
\bibitem{GeHa}M. Gell-Mann and J.\ Hartle , Phys. Rev. {\bf D 47}, 3345
(1993).
\bibitem{PaZu}J.\ B.\ Paz and W.\ H.\ Zurek, Phys. Rev. {\bf D 48} 2728
(1992).
\bibitem{SBP1}R.\ Schack, T.\ A.\ Brun, and I.\ Percival, J. Phys. {\bf A
28}, 5401 (1995).
\bibitem{SBP2}R.\ Schack, T.\ A.\ Brun, and I.\ Percival,  Phys. Rev. {\bf
A 53},
2694 (1996).


\end{thebibliography}
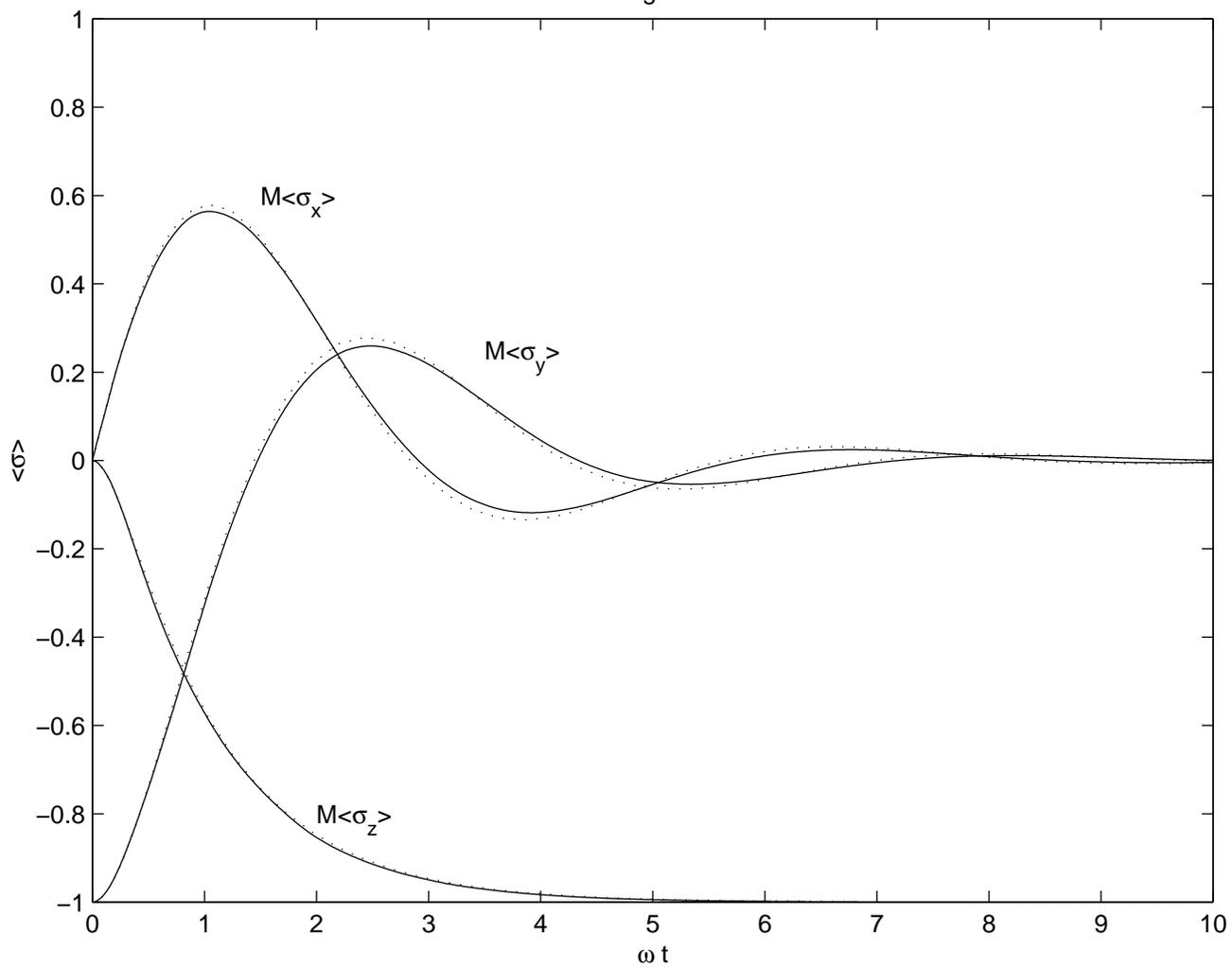
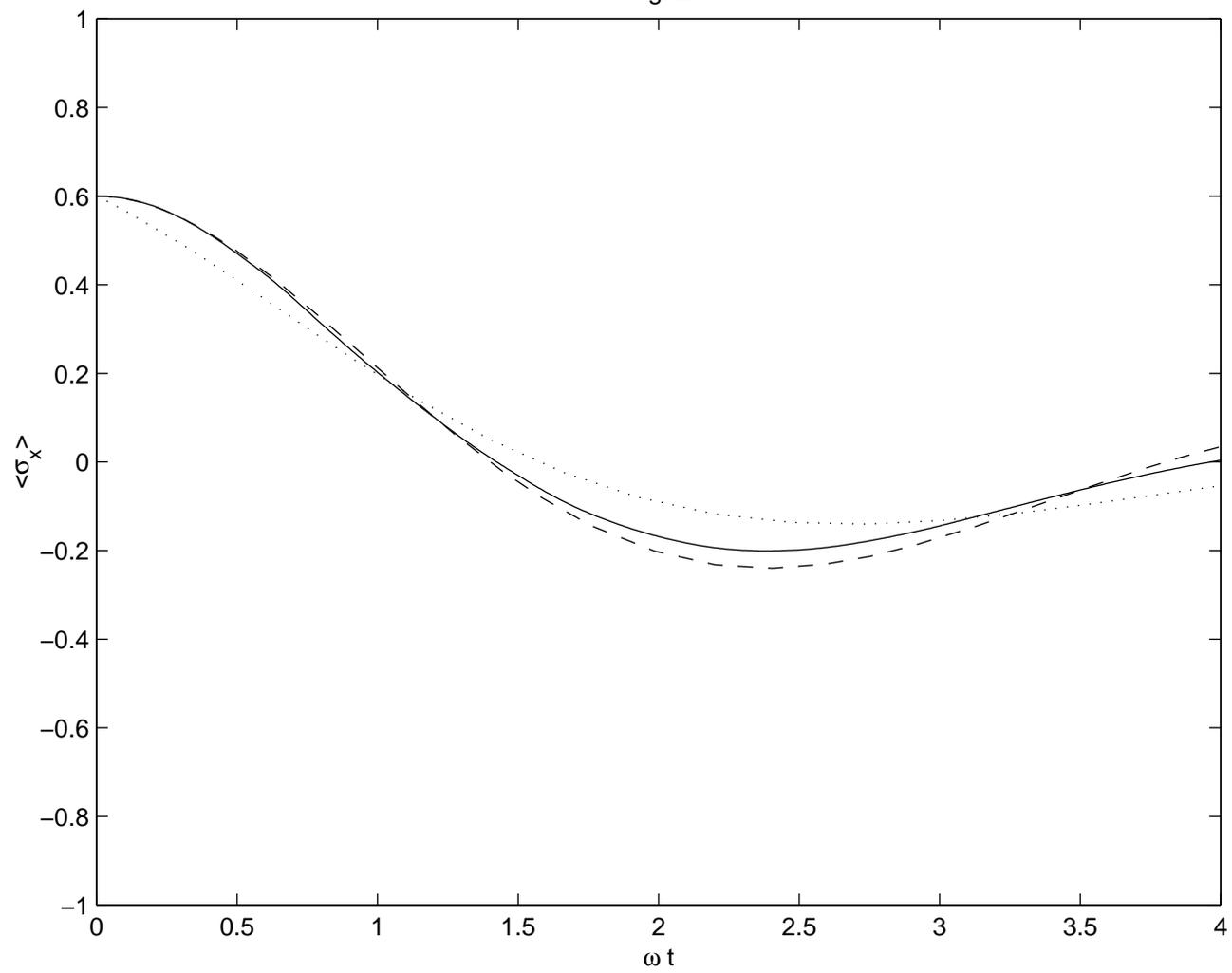
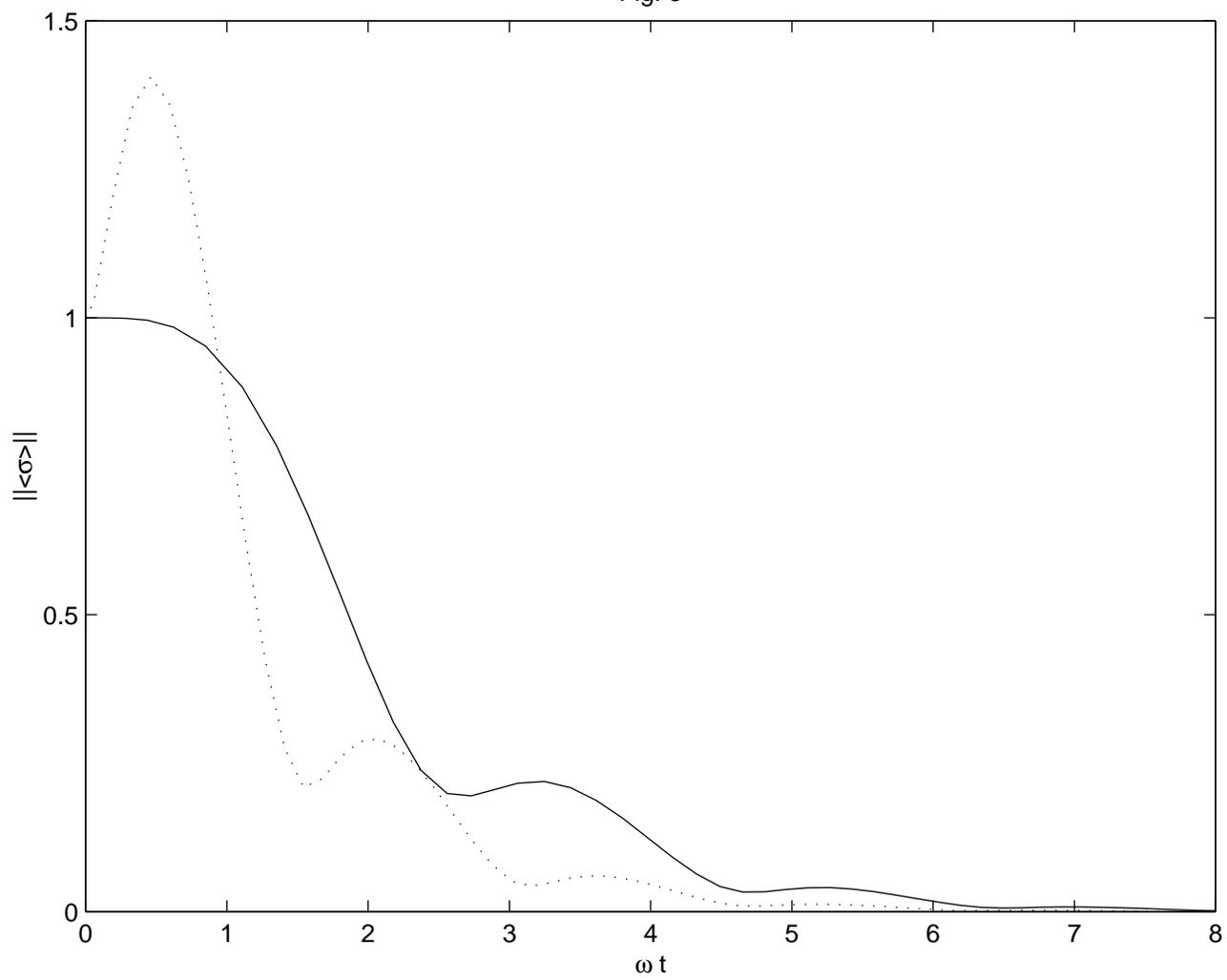
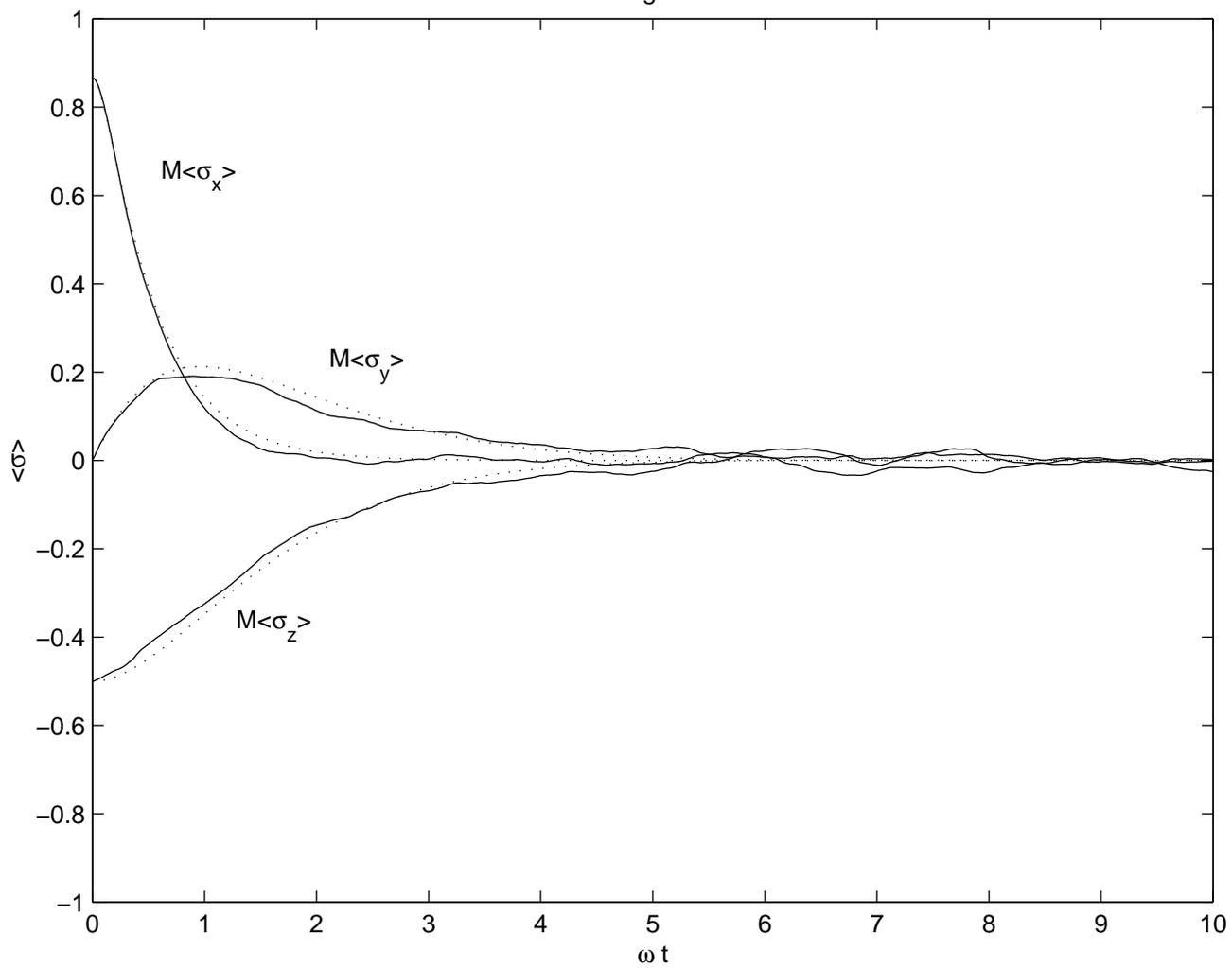
\end{document}